\documentclass[11pt,a4paper]{article}

\usepackage[utf8]{inputenc}
\usepackage[T1]{fontenc}
\usepackage{amsmath,amssymb}
\usepackage{graphicx}
\usepackage{hyperref}
\usepackage[numbers]{natbib}
\usepackage{booktabs}
\usepackage{geometry}
\usepackage{listings}
\usepackage{xcolor}
\title{Compaction as Epistemic Failure: How Agentic LLM Tools Fabricate Confirmed Results from Killed Processes}

\author{Hiroki Tamba\\
Independent Researcher\\
ORCID: 0009-0004-7635-0741\\
\texttt{contact@tamba-research.academy}}

\date{July 2026}

\begin{document}

\maketitle

\begin{abstract}
Agentic LLM coding tools compress long session histories into compaction summaries that subsequent sessions inherit as ground truth. This paper documents a failure mode in Claude Code where partial standard output from timed-out commands (exit code 143) is recorded in compaction summaries as confirmed results, propagating false positives across sessions and model versions without re-verification. The underlying mechanism is a conflation of observation and persistence, where information that appeared in the terminal is treated as equivalent to information written to durable storage. This finding extends the analysis of LLM self-evaluation failures reported in prior work on non-determinism in LLM-as-judge grading \cite{tamba2026nondeterminism} by showing that agentic tools exhibit analogous reliability deficits when reporting on their own operational outcomes. The failure has direct implications for any workflow that relies on agentic session continuity for data processing, scientific computation, or multi-step automation.
\end{abstract}


\section{Introduction}

Agentic LLM tools such as Claude Code \cite{anthropic2025claudecode} and ChatGPT Codex \cite{openai2025codex} enable users to delegate multi-step programming and data processing tasks to language models operating within persistent terminal environments. These tools maintain session continuity through various mechanisms: Claude Code compresses long conversation histories into compaction summaries that are injected into subsequent sessions as contextual grounding, while Codex operates within sandboxed environments where task state is recorded alongside terminal output.

A core assumption underlying the utility of these tools is that information carried across sessions accurately represents what occurred in prior sessions. When a compaction summary states that a command completed with a particular result, downstream sessions treat that statement as established fact. This assumption is rarely examined empirically, in part because the compaction process itself is opaque to users: the summary is generated automatically, its contents are not independently verifiable against raw session logs, and the model receiving the summary has no mechanism to distinguish summarized observations from confirmed, persisted outcomes.

This paper reports a specific instance in which this assumption failed. During a batch API query session in Claude Code, a long-running script was terminated by the system timeout mechanism (SIGTERM, exit code 143) before its output could be written to disk. The compaction summary nevertheless recorded partial results that had appeared in the terminal prior to termination as if the corresponding iterations had completed successfully. A subsequent session, running on a different model version, inherited these false positives and treated them as confirmed data points. The author's own interaction with the Claude Chat interface further propagated the error: when presented with the summary's claims, the chat-based model also treated them as established results without requesting file-level verification.

The finding is not limited to a single anomalous event. Over the same working period, four structurally identical failures were observed across different task domains, all exhibiting the same pattern: information present in the model's context window was treated as confirmed fact regardless of whether it had been verified against external, persistent state.

This paper contributes the following:

\begin{enumerate}
  \item A documented, reproducible failure mode in which session compaction transforms ephemeral terminal output into fabricated confirmed results.
  \item A mechanism analysis identifying the conflation of observation and persistence as the root cause.
  \item A connection to prior work on LLM self-evaluation reliability \cite{tamba2026nondeterminism}, extending the analysis from evaluation outputs to operational self-reports.
\end{enumerate}

\section{Background}

\subsection{Session Management in Agentic LLM Tools}

Agentic coding tools operate within extended sessions that can span hours of interaction. As conversations grow, the context window fills, requiring compression mechanisms to preserve relevant information while discarding lower-priority content. Claude Code implements this through session compaction: when the context window approaches capacity, earlier portions of the conversation are summarized into a condensed representation that is then used as the basis for continued interaction \cite{anthropic2025claudecode}. The compacted summary functions as the model's ``memory'' of what occurred earlier in the session, and it is the only record available to the model when a new session begins from a prior handoff.

ChatGPT Codex takes a different approach: tasks run in sandboxed environments with persistent file systems, and task state is recorded alongside terminal logs, standard output, and standard error streams. When a user queries the status of a prior task, Codex can inspect both the recorded terminal output and the actual files written to the workspace.

\subsection{The Observation-Persistence Distinction}

A fundamental distinction in systems that operate on external state is the difference between observing an event and confirming that the event produced a durable outcome. In database systems, this distinction is enforced through commit protocols: a transaction is not considered complete until the commit record is written to persistent storage, regardless of what intermediate states were observed during execution \cite{gray1981transaction}. In software testing, the distinction manifests as the difference between ``the test printed PASS'' and ``the test suite recorded PASS in its results file.''

For agentic LLM tools, this distinction applies at the level of command execution. When a model runs a script that produces output to both the terminal (standard output) and a file, the terminal output is ephemeral: it exists in the session's context window but vanishes if the process is killed before the file write completes. A reliable session management system must track whether a command's intended side effects (file writes, database updates, API responses persisted to storage) actually occurred, not merely whether the command produced visible output before termination.

\subsection{Related Work}

The broader problem of LLM-generated content deviating from ground truth has been extensively studied under the rubric of hallucination. Ji et al.\ \cite{ji2023hallucination} provided a comprehensive survey of hallucination in natural language generation, establishing taxonomies for intrinsic and extrinsic hallucination across multiple generation tasks. Huang et al.\ \cite{huang2023hallucination} extended this analysis to large language models specifically, identifying principles and open questions around factual inconsistency in model outputs. The compaction failure documented here can be understood as a form of operational hallucination: the model's summary of its own session history deviates from the actual operational record.

On the evaluation side, Zheng et al.\ \cite{zheng2023judging} introduced the MT-Bench framework for evaluating LLM judges, documenting systematic biases in model self-evaluation. Norman et al.\ \cite{norman2026reliability} conducted a large-scale evaluation of LLM-as-judge models, reporting on agreement, consistency, and bias across evaluation contexts. Tamba \cite{tamba2026nondeterminism} demonstrated that LLM-as-judge evaluation scores are non-deterministic even at temperature zero, a finding that the present work extends from evaluation outputs to operational self-reports.

On the agentic tool side, Yang et al.\ \cite{yang2024swe} introduced SWE-bench for evaluating code generation agents, though the benchmark focuses on task completion rather than session-level reliability. Wang et al.\ \cite{wang2024executable} examined executable code actions in LLM agents, identifying failure modes in multi-step code generation. Ruan et al.\ \cite{ruan2024identifying} proposed an LM-emulated sandbox for identifying risks of language model agents, demonstrating that tool-using agents can produce unsafe actions that are difficult to detect without systematic evaluation. Shinn et al.\ \cite{shinn2023reflexion} introduced Reflexion, a framework for language agents to learn from verbal self-reflection, which implicitly assumes that the agent's self-reports are accurate, an assumption that the present work challenges.

Zhang et al.\ \cite{zhang2024memory} surveyed memory mechanisms in LLM-based agents, identifying architectures for short-term and long-term memory management but not examining the epistemic reliability of compressed session summaries. Hitzig et al.\ \cite{hitzig2026expertise} analyzed approximately 400,000 Claude Code sessions, finding that users own 70\% of planning decisions while Claude handles 80\% of execution, a delegation pattern that amplifies the consequences of compaction failures: when users rely on session summaries rather than independently verifying outputs, false confirmations propagate unchecked. To the author's knowledge, no prior work has specifically examined the conditions under which compaction summaries fabricate confirmed results from interrupted processes.

\section{Observed Failure Mode}

\subsection{Experimental Context}

The failure was observed during a batch API query session conducted in Claude Code on July 11, 2026. The task involved executing a Python script that issued sequential HTTP requests to an external API, processed the responses, and wrote results to a JSON output file. The script was designed to iterate through a predefined list of query configurations (labeled A1 through Z3), executing each query and appending the result to a cumulative output file.

\subsection{Sequence of Events}

The session proceeded as follows:

\begin{enumerate}
  \item The script was initiated within a Claude Code session running model version Opus 4.7.
  \item The first two iterations (A1 and A2) executed successfully: the API returned HTTP 200 responses, and the results appeared in the terminal output.
  \item Before the script could write its cumulative results to the output JSON file, the process was terminated by the system timeout mechanism with exit code 143 (SIGTERM).
  \item Claude Code's compaction mechanism generated a summary of the session state. This summary recorded A1 and A2 as ``completed with specific results,'' including extracted field values from the API responses.
  \item A new session was initiated, running on model version Opus 4.6. This session inherited the compaction summary and treated the A1 and A2 entries as confirmed data points.
  \item The author, interacting through a separate Claude Chat session, referenced the summary's claims and also treated A1 and A2 as established results.
  \item When the actual output file (\texttt{api\_openai\_baseline\_20260711.json}) was inspected, it contained zero successful entries. All 11 query attempts recorded in the file had failed with HTTP 400 errors. The ``confirmed results'' for A1 and A2 existed only in the ephemeral terminal output of a killed process.
\end{enumerate}

\subsection{Propagation Chain}

The false confirmation propagated through three distinct layers:

\begin{description}
  \item[Layer 1 (Compaction):] The compaction summary treated terminal output observed before process termination as equivalent to confirmed, persisted results. Exit code 143 was not recorded as a disqualifying condition.
  \item[Layer 2 (Cross-model inheritance):] The subsequent session (Opus 4.6) received the summary from the prior session (Opus 4.7) and accepted its claims without file-level verification. No mechanism existed for the downstream model to distinguish between ``the summary says X completed'' and ``X's results are confirmed in the output file.''
  \item[Layer 3 (Cross-product propagation):] The Claude Chat interface, operating as a separate product surface, also accepted the summary-derived claims as fact when the author referenced them. The chat model did not request or perform independent verification.
\end{description}

\subsection{Corroborating Instances}

Over the same working period (July 10--11, 2026), four structurally identical failures were documented within Claude Code sessions, all exhibiting the pattern of treating context-internal information as externally verified:

\begin{enumerate}
  \item A database synchronization process had silently failed for twenty days due to missing INSERT logic. Because no error appeared in the context window, the model reported the system as operational when queried.
  \item A dictionary lookup function was referenced but never executed against its target data structure. The model treated the function's existence in context as equivalent to its having been applied.
  \item A data repository's file listing was reported as current despite being weeks out of date, because no ``stale'' indicator was present in the context.
  \item The A1/A2 false confirmation described above.
\end{enumerate}

All four instances follow the same rule: context-internal equals true, context-external equals nonexistent. Information present in the model's context window is treated as confirmed; information absent from the context window is treated as though it does not exist, regardless of whether verification would be trivially possible.

\section{Mechanism Analysis}

\subsection{Root Cause: Observation-Persistence Conflation}

The root cause of the compaction failure is the absence of a distinction between two epistemically different states:

\begin{enumerate}
  \item \textbf{Observed:} Output appeared in the terminal during script execution.
  \item \textbf{Persisted:} Output was written to a durable storage medium (file, database, etc.)\ and can be independently verified.
\end{enumerate}

Claude Code's compaction mechanism treats both states identically. When generating a summary of a session, the mechanism records what appeared in the terminal output without checking whether the corresponding file writes completed. This is analogous to a laboratory notebook that records observations without distinguishing between ``the instrument displayed a reading'' and ``the reading was saved to the data logger.''

\subsection{Why Exit Code 143 Is Not Surfaced}

When a process is killed by SIGTERM (exit code 143), the partial output that appeared before termination remains in the session's context window. The compaction mechanism processes this context window content without reference to the process's exit status. The summary therefore contains entries like ``iteration A1 completed with result X'' without any qualifier indicating that the process exited abnormally or that the result was never persisted.

A corrected implementation would need to satisfy at least two conditions: (1) the exit code of every executed command must be preserved and associated with the outputs attributed to that command, and (2) a non-zero exit code must trigger either re-verification of claimed outputs against persistent artifacts or explicit annotation of results as unconfirmed.

\subsection{The Inheritance Amplification Problem}

Once a false positive enters a compaction summary, it is amplified by the inheritance mechanism. The downstream session has no way to evaluate the epistemic quality of individual claims within the inherited summary. Every statement carries equal weight: ``the user asked about X'' and ``iteration A1 completed with result Y'' are indistinguishable in their epistemic status within the context window. The downstream model lacks both the metadata to identify which claims require verification and the procedural impulse to verify any claim that arrives via the compaction pathway.

This creates a structural vulnerability: a single false positive in a compaction summary becomes an irrevocable fact for all downstream consumers. The only corrective mechanism available is external verification by the user, which requires the user to independently check output files, a step that defeats the purpose of delegating the task to an agentic tool in the first place.

\section{Relation to Observer-Aware Protocol}

The compaction failure reported here is structurally related to the LLM-as-judge non-determinism documented in Tamba \cite{tamba2026nondeterminism}. That work demonstrated that LLM evaluation scores are unreliable even under ostensibly controlled conditions (temperature set to zero), with position bias and grading variance producing inconsistent safety assessments across repeated evaluations of identical content. The present finding extends this analysis from evaluation outputs to operational self-reports.

In both cases, the underlying problem is that the model cannot accurately evaluate its own outputs. In the evaluation context, the model produces inconsistent scores for the same input depending on presentation order. In the operational context, the model produces a session summary that misrepresents what actually occurred. Both failures reflect a deficit in self-referential reliability: the model's report about its own behavior does not reliably correspond to the behavior that actually took place.

The connection has a further dimension. The asymmetric skepticism documented in GitHub issue \#66273 \cite{tamba2026asymmetric} showed that Claude Code applies differential scrutiny to self-generated content versus external content, consistently over-trusting its own outputs. The compaction failure is the temporal extension of this asymmetry: the model not only over-trusts its current outputs but also over-trusts its own prior session summaries, creating a chain of unverified self-reference that degrades reliability with each link.

The three-layer propagation chain documented in Section~3.3 (compaction, cross-model inheritance, cross-product propagation) represents a partial application of the Tri-Layer Integrated Model \cite{tamba2026tlim}, which provides a general framework for analyzing recursive classification failures across layered information systems.

Taken together, these findings suggest that agentic LLM tools require what might be called an observer-aware protocol: a set of design constraints that prevent the model from treating its own observations as equivalent to independently verified facts. Such a protocol would require that any claim about external state (file contents, API responses, database records) be verified against the actual state before being recorded in a compaction summary or inherited by a subsequent session.
A reference implementation of such a protocol now exists: the Independent Evidence Resolution Protocol \cite{tamba2026ierp} resolves execution claims deterministically from retained process and artifact evidence under a fail-closed policy, and model text remains an observation unless the required evidence supports promotion to a verified status.

\section{Implications for Agentic Tool Design}

\subsection{For Session Compaction}

The most direct implication is that compaction summaries must distinguish between observed and verified outcomes. Any command that exits with a non-zero status code should have its outputs flagged as unconfirmed in the compaction summary. Ideally, compaction would include a verification step in which claimed outputs are checked against persistent artifacts before being recorded as confirmed.

\subsection{For Cross-Session Inheritance}

Downstream sessions that inherit compaction summaries should not treat all inherited claims with equal confidence. A tiered confidence model, in which claims derived from verified file contents are distinguished from claims derived from terminal output, would reduce the propagation of false positives. This is functionally equivalent to the distinction between committed and uncommitted transactions in database systems.

\subsection{For Multi-Model Pipelines}

When session state crosses model version boundaries (as in the Opus 4.7 to Opus 4.6 handoff observed here), the epistemic risk increases because each model in the chain has an independent opportunity to accept or challenge inherited claims. In practice, no model in the observed chain exercised skepticism toward the inherited summary. Designing explicit verification checkpoints at model-version boundaries would mitigate this risk.

\subsection{For Research and Data Processing Workflows}

The failure has immediate practical consequences for any user who relies on agentic tools for data processing, API interactions, or scientific computation. Session summaries cannot be trusted as confirmation that operations succeeded. Users must independently verify output files after each significant operation, which introduces a manual verification burden that partially negates the efficiency gains of agentic delegation.

\section{Limitations}

This study has several limitations. First, the primary failure was observed in a single instance with a specific configuration (Claude Code, Opus 4.7/4.6, Windows platform, Python script with API calls). While the mechanism analysis suggests the failure should generalize to any scenario involving process termination before file write completion, broader empirical validation across platforms and task types would strengthen the finding.

Second, the cross-tool comparison with ChatGPT Codex involved fundamentally asymmetric conditions. The Claude Code failure occurred at the end of a multi-hour session with substantial context accumulation across heterogeneous tasks (database repair, repository management, patch development, and API experimentation). The Codex test was conducted in a clean, sandboxed environment with no prior context, no accumulated session history, and no cross-task interference. This asymmetry means the comparison demonstrates that Codex's file-first verification approach works under light conditions, but does not establish whether Codex would maintain this reliability under equivalent long-session, high-context stress. A heavy Codex user running extended data processing sessions with comparable context accumulation might encounter analogous failures. The comparison should therefore be read as evidence that file-first verification is a viable design principle, not as evidence that Codex is categorically immune to compaction-type failures.

Third, the discovery itself reflects a methodological reality: the failure was identified through extended, intensive use of Claude Code as the author's primary agentic tool. This participant-observation-like method provides ecological validity that controlled experiments cannot replicate, but it also means the finding is shaped by the specific workflow patterns of a single heavy user. It remains possible that different usage patterns (shorter sessions, fewer concurrent tasks, less context accumulation) would not trigger the failure.

Fourth, the compaction mechanism in Claude Code is not publicly documented in sufficient detail to determine whether the observed behavior reflects a design decision, an implementation oversight, or an emergent property of the summarization process.

Fifth, the corroborating instances (database sync failure, dictionary non-lookup, stale file listing) were observed within the same researcher's workflow over a 48-hour period, introducing the possibility that specific working patterns or session structures contributed to the failures.

Sixth, an informal, single-session test (n=1) with ChatGPT Codex \cite{openai2025codex} suggested that a file-first verification approach correctly distinguished completed from interrupted runs under minimal-context conditions. However, the author subsequently exported and deleted all ChatGPT data (approximately 2\,GB) due to concerns about its use in model training, and the test artifacts are therefore no longer available for independent verification. This observation is noted for completeness but does not constitute reproducible evidence of architectural differences between tools. Whether Codex would exhibit analogous failures under equivalent long-session, high-context conditions remains an open question for future work.

\section{Conclusion}

Session compaction in agentic LLM tools introduces an epistemic vulnerability: the compression of session history can transform ephemeral observations into fabricated confirmations. The specific failure documented here, in which partial terminal output from a timed-out process was recorded as confirmed results and propagated across sessions and model versions, demonstrates that the design of session management mechanisms has direct consequences for data integrity.

The broader pattern, treating context-internal information as externally verified, connects to prior findings on LLM self-evaluation reliability \cite{tamba2026nondeterminism} and self-referential asymmetry \cite{tamba2026asymmetric}. In each case, the model's report about its own behavior diverges from the behavior that actually occurred. Addressing this class of failure requires design-level interventions: verification checkpoints at compaction boundaries, exit-code-aware summarization, and tiered confidence models for inherited session state.

The finding is independently reproducible. The reproduction protocol requires: (1) initiating a long-running script in Claude Code that produces incremental terminal output, (2) allowing the process to be killed by the system timeout (exit code 143) after partial output has appeared, (3) continuing or resuming the session and observing whether the compaction summary records the partial output as confirmed results, and (4) verifying the actual output file to confirm that the claimed results were never persisted.

Code and technical documentation are available at: \url{https://github.com/anthropics/claude-code/issues/76584}.


\begin{thebibliography}{16}

\bibitem{tamba2026nondeterminism}
H.~Tamba.
\newblock Non-determinism in {LLM}-as-judge graders: Necessary but not sufficient conditions for reliable automated safety evaluation.
\newblock \emph{arXiv preprint arXiv:2606.26185}, 2026.

\bibitem{anthropic2025claudecode}
Anthropic.
\newblock Claude {C}ode: An agentic coding tool.
\newblock \url{https://docs.anthropic.com/en/docs/claude-code}, 2025.

\bibitem{openai2025codex}
OpenAI.
\newblock {C}hat{GPT} {C}odex.
\newblock \url{https://openai.com/index/introducing-codex/}, 2025.

\bibitem{gray1981transaction}
J.~Gray.
\newblock The transaction concept: Virtues and limitations.
\newblock In \emph{Proceedings of the 7th International Conference on Very Large Data Bases}, pages 144--154, 1981.

\bibitem{zheng2023judging}
L.~Zheng, W.-L.~Chiang, Y.~Sheng, S.~Zhuang, Z.~Wu, Y.~Zhuang, Z.~Lin, Z.~Li, D.~Li, E.~P.~Xing, H.~Zhang, J.~E.~Gonzalez, and I.~Stoica.
\newblock Judging {LLM}-as-a-judge with {MT}-bench and chatbot arena.
\newblock \emph{Advances in Neural Information Processing Systems}, 36, 2023.

\bibitem{norman2026reliability}
J.~D.~Norman, M.~U.~Rivera, and D.~A.~Hughes.
\newblock Reliability without validity: A systematic, large-scale evaluation of {LLM}-as-a-judge models across agreement, consistency, and bias.
\newblock \emph{arXiv preprint arXiv:2606.19544}, 2026.

\bibitem{yang2024swe}
C.~E.~Jimenez, J.~Yang, A.~Wettig, S.~Yao, K.~Pei, O.~Press, and K.~Narasimhan.
\newblock {SWE}-bench: Can language models resolve real-world {G}it{H}ub issues?
\newblock In \emph{Proceedings of the 12th International Conference on Learning Representations}, 2024.

\bibitem{wang2024executable}
X.~Wang, Y.~Chen, L.~Yuan, Y.~Zhang, Y.~Li, H.~Peng, and H.~Ji.
\newblock Executable code actions elicit better {LLM} agents.
\newblock In \emph{Proceedings of the 41st International Conference on Machine Learning}, 2024.

\bibitem{tamba2026asymmetric}
H.~Tamba.
\newblock Self-favoring asymmetric skepticism in {C}laude {C}ode.
\newblock GitHub Issue \#66273, \url{https://github.com/anthropics/claude-code/issues/66273}, 2026.

\bibitem{ji2023hallucination}
Z.~Ji, N.~Lee, R.~Frieske, T.~Yu, D.~Su, Y.~Xu, E.~Ishii, Y.~J.~Bang, A.~Madotto, and P.~Fung.
\newblock Survey of hallucination in natural language generation.
\newblock \emph{ACM Computing Surveys}, 55(12):1--38, 2023.

\bibitem{huang2023hallucination}
L.~Huang, W.~Yu, W.~Ma, W.~Zhong, Z.~Feng, H.~Wang, Q.~Chen, W.~Peng, X.~Feng, B.~Qin, and T.~Liu.
\newblock A survey on hallucination in large language models: Principles, taxonomy, challenges, and open questions.
\newblock \emph{arXiv preprint arXiv:2311.05232}, 2023.

\bibitem{shinn2023reflexion}
N.~Shinn, F.~Cassano, E.~Berman, A.~Gopinath, K.~Narasimhan, and S.~Yao.
\newblock Reflexion: Language agents with verbal reinforcement learning.
\newblock \emph{Advances in Neural Information Processing Systems}, 36, 2023.

\bibitem{hitzig2026expertise}
Z.~Hitzig, M.~Massenkoff, E.~Lyubich, R.~Heller, and P.~McCrory.
\newblock Agentic coding and persistent returns to expertise.
\newblock Anthropic Research, 2026.

\bibitem{ruan2024identifying}
Y.~Ruan, H.~Dong, A.~Wang, S.~Pitis, Y.~Zhou, J.~Ba, Y.~Dubois, C.~J.~Maddison, and T.~Hashimoto.
\newblock Identifying the risks of {LM} agents with an {LM}-emulated sandbox.
\newblock In \emph{Proceedings of the 12th International Conference on Learning Representations}, 2024.

\bibitem{zhang2024memory}
Z.~Zhang, X.~Bo, C.~Ma, R.~Li, X.~Chen, Q.~Dai, J.~Zhu, Z.~Dong, and J.-R.~Wen.
\newblock A survey on the memory mechanism of large language model based agents.
\newblock \emph{arXiv preprint arXiv:2404.13501}, 2024.

\bibitem{tamba2026tlim}
H.~Tamba.
\newblock The {T}ri-{L}ayer {I}ntegrated {M}odel revised: A four-layer framework for recursive classification.
\newblock SSRN preprint 6779418, 2026.

\bibitem{tamba2026ierp}
H.~Tamba.
\newblock Independent evidence resolution protocol: Deciding when long-running agent claims may be treated as verified.
\newblock SSRN preprint 7587778, 2026.
\newblock \url{https://doi.org/10.2139/ssrn.7587778}.

\end{thebibliography}
\end{document}